\documentclass{aastex}
\usepackage{spr-astr-addons}
\usepackage{url}

\RequirePackage{color}
\shorttitle{Removing the Big Bang Singularity}
\shortauthors{R. Rashidi}

\begin{document}

\title{Removing the Big Bang Singularity: The role of the generalized uncertainty principle in quantum gravity}

\author{Reza Rashidi}
\affil{Department of Physics, Shahid
Rajaee Teacher Training University, Lavizan, Tehran 16788, Iran}
\email{reza.rashidi@srttu.edu}

\begin{abstract}
The possibility of avoiding the big bang singularity by means of a generalized uncertainty principle is investigated. In relation with this matter, the statistical mechanics of a free-particle system obeying the generalized uncertainty principle is studied and it is shown that the entropy of the system has a finite value in the infinite temperature limit. It is then argued that negative temperatures and negative pressures are possible in this system. Finally, it is shown that this model can remove the big bang singularity.
\end{abstract}


\section{Introduction}
The big bang singularity is a long standing issue appearing in the standard theory of cosmology as well as inflation theory \citep{Hawking73,Borde94}. Since the tensors involved in the Einstein equations there, diverge, it is widely believed that the prediction of such a singularity within the framework of the general theory of relativity is a signal that the physical theory is not valid at this extreme condition. In fact, it is expected that by approaching the big bang singularity the quantum effects become important and it becomes necessary to apply a quantum theory of gravity. Much effort has been made in the framework of quantum theories of gravity to solve the problem of singularity. Scenarios from string theory and loop quantum gravity being some of them \citep{Kawai06,Gasperini93,Larsen97, Bojowald01,Ashtekar03}.
A quantum theory of gravity is originated from the unification of general relativity and
quantum mechanics. This unification can lead to a fundamental minimal length \citep{Garay95,Jaeckel94}, effect of which, in quantum mechanics, might be described as a non-zero
minimal uncertainty $\triangle x_0$ in position measurements. For
example, string theory leads to a minimal length which is effectively of the
form of a minimal position uncertainty \citep{Amati89,Maggiore93a,Maggiore93b}. In this paper we are going to investigate whether the appearance of a nonzero minimal uncertainty in position as an effect of quantum gravity can remove the big bang singularity.

To generalize the uncertainty relation which implies the appearance
of a nonzero minimal uncertainty $\triangle x_0$ in position, one
has to modify the commutation relation between the position
operator and the momentum operator. Such generalizations are not
unique. For example, one can take the commutation relation as
\begin{equation}
[X, P]=\imath\hbar(1+\lambda^2P^2) \label{1}
\end{equation}
(known as the Kempf-Mangano-Mann (K.M.M.) deformation \citep{Kempf95}). Maggiore's generalization is another example as following \citep{Maggiore94}
\begin{equation}
[X, P]=\imath\hbar\sqrt{1+\lambda^2(P^2+m^2c^2)} \label{2}
\end{equation}
where $\lambda$ is a very small length parameter. The first
commutation relation yields the generalization uncertainty relation
\begin{equation}
\triangle x\triangle p\geq\hbar(1+\lambda^2\triangle p ^2) \label{3}
\end{equation}
and the second relation also leads to (\ref{3}) in an appropriate limit.
A natural generalization of (\ref{1}) in three dimensions which preserves the rotational symmetry is \citep{Kempf95}
\begin{equation}
[X_{i}, P_{j}]=\imath\hbar\delta_{ij}(1+\lambda^2\bar{P}^2)
\end{equation}
\begin{equation}
[P_{i}, P_{j}]=0
\end{equation}
\begin{equation}
[X_{i}, X_{j}]=2\imath\hbar\lambda^2(P_{i}X_{j}-P_{j}X_{i}). \label{3b}
\end{equation}
The last relation leads to a noncommutative geometry.

The statistical mechanics of free particle
systems obeying the generalized uncertainty principle leads to many
novel consequences \citep{Brout99, Lubo00a, Maggiore94}. Rama has studied the statistical mechanics of
Maggiore's generalization in the grand canonical ensemble approach \citep{Rama01}. He has shown that there is a drastic reduction in the degrees of freedom in the high temperature limit. Some aspects
of statistical mechanics of K.M.M. generalization have also been studied
in the canonical ensemble approach \citep{Lubo00b}.

In this paper we first study the statistical mechanics of the
systems of free ultra-relativistic particles obeying K.M.M.
deformation in the grand canonical ensemble approach. We introduce
some new statistical aspects of this generalization.
The equation of state is obtained and it will be shown that, in
contrast to Maggiore's generalization and conventional quantum
statistical mechanics, the entropy and the internal energy of the system have finite values
in the infinite temperature limit. Then it will be argued that these results are actually the consequences of the fact that the total number of accessible states for such a system is finite. There is another well known system which behaves similarly. A spin system such as a paramagnetic system has a finite number of states and hence the entropy and the internal energy of the system take finite values as the temperature tends to infinity \citep{Greiner95,Pathria96}. Another interesting property of spin systems is possibility of negative temperatures. In fact, the existence of negative temperature is an integral part of a system with a finite number of states \citep{Ramsey56}. Then we argue that it is similarly possible to have negative temperatures and, moreover, negative pressures in a system of free particles obeying the generalized uncertainty principle (\ref{3}). It is important to note that there is a difference between such a system and spin systems. In spin systems the
translational degrees of freedom of a system (i.e., coordinates and momenta) are not taken into account, but only its spin degrees of freedom are considered. But in our system we do not eliminate such degrees of freedom and this is the reason why negative pressures occur in such systems and do not appear for spin systems. We also
obtain the equation of state in the negative temperature region. Finally, we investigate the consequence of the equation of state in the Friedmann equations and show that this model can remove the big bang
singularity.

\section{The statistical mechanics}

In this section we study the statistical mechanics of a system of
free ultra-relativistic particles confined in a 3-dimensional volume
$V$ obeying the generalized uncertainty principle (\ref{3}) by
means of the grand canonical ensemble.

In conventional quantum statistical mechanics where $\lambda=0$,
due to the Heisenberg uncertainty principle, the phase space is
divided into cells of volume $h^3$ where $h$ is the Planck
constant. This means that the one-particle phase space measure is
given by $h^{-3}$ and therefore each integral on the phase space will be of
the form $\int d^3x d^3p h^{-3}(\Box)$. Consequently, as it has been shown in \citep{Chang02} for particles
obeying the generalized uncertainty principle (\ref{3}), the phase
space should be divided in to cells of volume $h^{3}(1+\lambda^2
p^2)^3$ and thus phase space integrals are of the form $\int d^3x d^3p
h^{-3}(1+\lambda^2 p ^2)^{-3}(\Box)$.

To calculate the grand canonical partition function of a non-relativistic ideal gas in the conventional quantum mechanics, the sum over all one-particle states, for a large volume, can be rewritten in terms of an integral as
\begin{eqnarray}
\sum_{\vec{n}}\longrightarrow\frac{V}{(2\pi)^3}\int d^3k&=&\frac{2\pi V}{h^3}(2m)^{3/2}\int E^{1/2}dE\nonumber\\
&=&\int\frac{d^3x d^3p}{h^3},
\end{eqnarray}
where $\vec{k}=\frac{\pi}{L} (n_{x},n_{y},n_{z})$, $E=\frac{\hbar^2}{2m}\vec{k}^2$ and for the last integral $E=\vec{p}^2/2m$ was applied. This calculation shows that the phase-space measure $h^{-3}$ can be automatically obtained when we evaluate the large volume limit of the sum over quantum states \citep{Greiner95}.
Therefore, the aforementioned consideration, concerning the phase-space measure for particles obeying the generalized uncertainty principle, becomes more justifiable if we prove that the integral on the phase space with the measure $h^{-3}(1+\lambda^2 p ^2)^{-3}$ is the large volume limit of the sum over the quantum mechanical states.
To show this, for simplicity we consider a massive and non-relativistic particle obeying K.M.M. deformation in a one dimensional box. First note that in this case the definition of a box in $\hat{x}$-space is impossible because the position operator $\hat{x}$ is no longer self adjoint \citep{Kempf95} and then the representation of a potential corresponding to a box with exact positions for its walls in $\hat{x}$-space is not possible. From a physical point of view, it means that since we have a non-zero minimal uncertainty in position the construction of a box with the exact wall location in position space is impossible. But, as mentioned in \citep{Kempf95}, one can define a quasi-position space and consider a box in it. Now, solving the Schr\"{o}dinger equation with boundary conditions corresponding to the box leads to the dispersion relation
\begin{equation}
E(k)=\frac{1}{2m\lambda^2}\tan^2(\hbar\lambda k)\hspace{0.5 cm},\hspace{0.5 cm} k=\frac{\pi n}{L} \label{4a}
\end{equation}
where $n$ is an integer and $L$ is the length of the box \citep{Lubo00b, Kempf95}. To calculate the grand canonical partition function of an ideal gas we have to determine the sum over $n$ (all one-particle states). But for a large volume, since $\triangle k=\frac{\pi}{L}\triangle n$ becomes very small, one can substitute the sum over $n$ with an integral as follows
\begin{equation}
\sum_{n}\rightarrow \frac{L}{2\pi}\int dk=\frac{L}{h}\int\sqrt{\frac{2m}{E}}\frac{dE}{(1+2m\lambda^2E)}\hspace{.3 cm}.
\end{equation}
For the last integral the dispersion relation (\ref{4a}) has been used. Now, by substituting $p^2/2m=E$ in the last integral we obtain
\begin{equation}
\sum_{n}\rightarrow \frac{L}{2\pi}\int dk=\int\frac{dx dp}{h(1+\lambda^2p^2)}.
\end{equation}
As one can see, the integral measure $h^{-1}(1+\lambda^2p^2)^{-1}$ automatically appears after substituting the sum by the integral.

It is clear that the effect of a nonzero minimal uncertainty in
position will be considerable in the high energy/temperature limit.
Since in this limit the mass of particles and the forces between
them are negligible, we can consider such particles to be free ultra-relativistic particles in the high temperature limit. In the high temperature limit one can assume that the particles obey
Maxwell-Boltzmann statistics. Thus, we have
\begin{equation}
\ln \mathcal{Z}=\beta \mathcal{P}V=\int\frac{e^{-\beta E}d^3x d^3p
}{h^{3}(1+\lambda^2 p ^2)^{3}}  \label{4}
\end{equation}
where $\mathcal{Z}$ is the grand canonical partition function,
$\beta$ is the inverse temperature, $\mathcal{P}$ is the pressure,
and $E$ is the eigen-value of the Hamiltonian of a free
ultra-relativistic particle (we assume that the chemical potential $\mu$ is zero because
the mass of an ultra-relativistic particle is negligible as compared to its energy). Since the
Hamiltonian of a free particle is $H=\sqrt{P^2}$, its eigen-values
are of the form $E=\sqrt{p_{x}^2+p_{y}^2+p_{z}^2}$ where $p_{x}$,
$p_{y}$ and $p_{z}$ denote the eigen-values of the momentum
operator. Then equation (\ref{4}) becomes
\begin{equation}
\ln \mathcal{Z}=\mathcal{C}\int^{\infty}_{0}\frac{E^2 e^{-\beta E}dE
}{(1+\lambda^2 E ^2)^{3}}  \label{5}
\end{equation}
where $\mathcal{C}\equiv\frac{4\pi V}{h^3}$. To obtain the partition
function in a closed form it is needed to evaluate integrals of the form
\begin{equation}
 \textrm{I}_{m,n}(x,\alpha)=\int^{\infty}_{0}t^n(t^2+\alpha^2)^{-m/2}e^{-xt}dt. \label{6}
\end{equation}
It is easy to see that \citep{Rama01}
\begin{equation}
 \textrm{I}_{2k,n}(x,\alpha)=\frac{(-1)^{n+k-1}}{(k-1)!}(\frac{d}{dx})^n
(\frac{d}{d\alpha^2})^{k-1}\textrm{I}_{2,0}(x,\alpha),
\end{equation}
where
\begin{equation}
 \textrm{I}_{2,0}(x,\alpha)=\frac{1}{\alpha}(\textrm{ci}(\alpha x)\sin(\alpha x)-\textrm{si}(\alpha x)\cos(\alpha x))
\end{equation}
and
\begin{equation}
\textrm{ci}(x)=-\int_{x}^{\infty}\frac{\cos(t)}{t}dt
\end{equation}
\begin{equation}
\textrm{si}(x)=-\int_{x}^{\infty}\frac{\sin(t)}{t}dt.
\end{equation}
Setting $t=\lambda E$ and $x=\beta/\lambda$ in equation (\ref{5}),
it can be shown that
\begin{equation}
\ln
\mathcal{Z}=\frac{\mathcal{C}}{\lambda^3}\int_{0}^{\infty}t^2(t^2+1)^{-3}e^{-xt}dt
=\frac{\mathcal{C}}{\lambda^3}\textrm{I}_{6,2}(x,1).\label{7}
\end{equation}
Now we can calculate various thermodynamical quantities such as the internal energy
\begin{eqnarray}
U=-\frac{\partial\ln
\mathcal{Z}}{\partial\beta}|_{\textit{z},V}&=&\frac{\mathcal{C}}{\lambda^4}\int_{0}^{\infty}t^3(t^2+1)^{-3}e^{-xt}dt\nonumber
\\&=&\frac{\mathcal{C}}{\lambda^4}\textrm{I}_{6,3}(x,1),\label{8}
\end{eqnarray}
where $\textit{z}=e^{\beta\mu}$ is the fugacity. Thus, the entropy becomes
\begin{equation}
S=\beta(U+\mathcal{P}V-\mu
N)=\frac{\mathcal{C}}{\lambda^3}\{\textrm{I}_{6,2}(x,1)+x\textrm{I}_{6,3}(x,1)\}.\label{9}
\end{equation}

To investigate the behavior of these thermodynamical quantities in
the high temperature limit $x=\frac{\beta}{\lambda}\ll1$, it is
appropriate to expand them around $x=0$. The Taylor expansion of the
partition function is
\begin{equation}
\ln \mathcal{Z}(x\ll1)\simeq
\frac{\mathcal{C}}{\lambda^3}(\frac{\pi}{16}-\frac{1}{4}x+\frac{3\pi
}{32}x^2),\label{10}
\end{equation}
where terms of higher than second order are omitted. Then for
the internal energy and the entropy we have
\begin{equation}
U(x)\simeq \frac{\mathcal{C}}{\lambda^4}(\frac{1}{4}-\frac{3\pi
}{16}x),\label{11}
\end{equation}
and
\begin{equation}
S(x)\simeq
\frac{\mathcal{C}}{\lambda^3}(\frac{\pi}{16}-\frac{3\pi}{32}x^2).\label{12}
\end{equation}
One can easily see that in the infinite temperature limit the value
of the internal energy and the value of the entropy are finite. This is in contrast
to Maggiore's generalization and conventional quantum statistical
mechanics. Although in the case of Maggiore's generalization there
is a drastic reduction in the degrees of freedom, in the infinite
temperature limit the values of the internal energy and the entropy
tend to infinity but slower than that in the standard case \citep{Rama01}. From equations (\ref{11}) and (\ref{12}) we have
\begin{equation}
U|_{\beta=0_+}=\frac{\mathcal{C}}{4\lambda^4}\hspace{0.3
cm},\hspace{0.4 cm} S|_{\beta=0_+}=\frac{\mathcal{\pi
C}}{16\lambda^3}.\label{13}
\end{equation}
The physical origin of these finite values is that the volume of the
phase space cells $h^{3}(1+\lambda^2 p^2)^3$ grows faster than $p^2$ such that the number of available cells always remains finite.
Moreover, at $\beta=0_+$ the derivative of the entropy $S$ with respect to the internal energy $U$ is zero since
\begin{equation}
\frac{dS}{dU}|_{\beta=0_+}=(\frac{dS}{d\beta}\frac{d\beta}{dU})|_{\beta=0_+}=0.\label{13b}
\end{equation}
In all ordinary systems where one considers the conventional quantum relations and the kinetic energy of the particles as a component of the internal energy, there is no bound on the internal energy and on the entropy in high temperature limit. As mentioned earlier, similar situation may arise in a spin system where the focus is only on the spin degrees of freedom and the translational degrees of freedom are neglected. A spin system, such as a system of interacting nuclear spins and an ideal paramagnetic system, has upper bound to its allowed energy and its total number of states is finite. Therefore, it is easy to see that the internal energy and the entropy of such a system, in the presence of an external magnetic field, approach finite values as $T\rightarrow\infty$. Also, the derivative of the entropy with respect to the internal energy in high temperature limit goes to zero \citep{Greiner95,Pathria96,Reif85,Ramsey56}. However, It should be emphasized again that the above results are completely novel because there is an important difference between present case and a spin system. In contrast to spin systems, we don't eliminate the kinetic energy of the particles in this case. On the other hand, spin systems have another important property which is an integral part of such systems. Since the total number of states available to a spin system is finite, the number of possible states initially increases with increasing energy; but then it reaches a maximum and decreases again. Thus negative temperatures as well as positive temperatures become possible \citep{Greiner95,Reif85,Ramsey56}. Since the behavior of a spin system and a system of particles obeying the generalized uncertainty principle are very similar in the high temperature limit, the question arises whether such a system can possess a negative temperature. We will argue that the answer may be the affirmative.

From a thermodynamic point of view, a system can have a negative temperature whenever the entropy is not restricted to a monotonically increasing function of the internal energy. At any point for which the derivative of the entropy with respect to the internal energy becomes negative, the temperature is negative. This requirement occurs when the energy levels of the elements of the system have an upper bound. In other words, in order to have a negative temperature the total number of possible states must be finite.
But in the system of particles obeying the generalized uncertainty principle which is the subject of the study in this paper, there is apparently no bound on the energy levels. Nevertheless, we show that, from a practical point of view, the total number of states of such a system is finite and then it is reasonable to set a cutoff on energy because a state with a finite quantum number and an infinite energy eigen-value is not a physical state and it is legitimate to exclude it.
Before studying the problem, note that if such a cutoff on energy is large enough, our calculations will not be affected in the positive temperature region because the
integrand in (\ref{5}), even by omitting the Boltzmann factor $e^{-\beta E}$,
rapidly tends to zero at infinity and taking the upper limit of integration to infinity, hence, becomes irrelevant.\\
Now again, it is useful to investigate a non-interacting system of massive and non-relativistic particles obeying K.M.M. deformation in a one dimensional space at the beginning. As mentioned before, to calculate the one-particle canonical partition function for a large volume the sum over one-particle states can be substituted with an integral as following
\begin{equation}
Z(\beta,L,1)=\sum_{n} e^{-\beta E_n}=\frac{L}{2\pi}\int e^{-\beta E(k)}dk,
\end{equation}
where $n$, $E_n$, $k$ and $L$ denote the quantum number, the eigen-value corresponding to $n$, the wave number and the length of the box, respectively and equation (\ref{4a}) gives the relations between them. Now, if we allow the upper limit of the sum to tend to infinity, the wave number $k$ runs over the interval $(-\infty, +\infty)$ and it is easy to see that
\begin{equation}
Z(\beta,L,1)=\frac{L}{2\pi}\int_{-\infty}^{+\infty} e^{-\beta E(k)}dk=\infty,\label{14a}
\end{equation}
for every $\beta$. Although the partition function becomes an indefinite quantity, one can definitely determine various thermodynamical quantities. For example, to calculate the internal energy one can write
\begin{eqnarray}
 U(\beta,L,N)&=&N\lim_{R\rightarrow\infty}\frac{\frac{L}{2\pi}\int_{-R}^{+R} E(k)e^{-\beta E(k)}dk}{\frac{L}{2\pi}\int_{-R}^{+R} e^{-\beta E(k)}dk}\nonumber\\&=&N\frac{\frac{L}{2\pi}\int_{-\frac{\pi}{2\hbar\lambda}}^{+\frac{\pi}{2\hbar\lambda}} E(k)e^{-\beta E(k)}dk}{\frac{L}{2\pi}\int_{-\frac{\pi}{2\hbar\lambda}}^{+\frac{\pi}{2\hbar\lambda}} e^{-\beta E(k)}dk}<\infty,
\end{eqnarray}
where $N$ is the total number of particles and in the last term we have employed the fact that $E(k)$ is a periodic function of $k$ with period $\frac{\pi}{\hbar\lambda}$. Therefore, to avoid the infinity appearing in (\ref{14a}) one can redefine the partition function as
\begin{eqnarray}
Z(\beta,L,1)&=&\frac{L}{2\pi}\int_{-\frac{\pi}{2\hbar\lambda}}^{+\frac{\pi}{2\hbar\lambda}} e^{-\beta E(k)}dk\nonumber\\&=&\frac{L}{h}\int_{0}^{\infty}\sqrt{\frac{2m}{E}}\frac{e^{-\beta E}dE}{(1+2m\lambda^2E)} \hspace{.5 cm},\label{14b}
\end{eqnarray}
which is finite for all $\beta>0$. Then the internal energy becomes
\begin{equation}
U(\beta,L,N)=-N\frac{\partial\ln(Z)}{\partial\beta}|_{L}\hspace{.5 cm}.
\end{equation}
It means that the cutoff on the wave number not only eliminates the infinity of the partition function but also leaves the thermodynamical quantities unchanged. Without such a cutoff, the situation in the grand canonical ensemble approach  is worse because in that case both the partition function and the internal energy become infinity. In addition to the above approach, one can argue that since there is a non-zero limit to the precision of position measurements, the Fourier decomposition of the quasi-position wave function of physical states can not contain wavelength components smaller than $4\hbar\lambda$ which sets a cutoff on the wave number \citep{Kempf95}.

Consequently, from both physical and practical points of view, it is necessary to consider a cutoff on the wave number. In other words, to obtain physical results it is enough to consider a finite number of states. Furthermore, from equation (\ref{14b}) one can see that this finite set of states covers the whole range of energy from zero to infinity. Therefore, in calculation of partition functions, taking the integral over the whole range of energy, in fact, means a summation over a finite set of states. Now, since the quantum number $n$ is finite, the dispersion relation (\ref{4a}) yields finite values for $E_n$ providing that $\frac{L}{\hbar\lambda}$ is not an integer. But even if $\frac{L}{\hbar\lambda}$ belongs to the set of natural numbers, we should exclude the state corresponding to $E(n_{max})=\infty$ because this single state which has a finite quantum number and infinite energy is not accessible by finite energy and, therefore, is not a physical state. However, we have to note that for an isolated system the finiteness of the total number of accessible states is enough to have negative temperatures because in this case the energy of the system as a constraint has a fixed finite value and worrying about the state possessing infinite energy value is irrelevant. The cutoff on energy becomes important whenever we deal with a system which is coupled to a heat bath at temperature $T$ because of the Boltzmann factor $e^{-\beta E_n}$. The above reasons show that a non-interacting system of massive and non-relativistic particles obeying K.M.M. deformation in a one dimensional space is actually a system with a finite number of states and a bound on its energy levels and thus it can possess a negative temperature.

The situation in three dimensional space is similar and one can see that negative temperature can occur in a non-interacting system of ultra-relativistic particles obeying K.M.M. deformation. To show this we consider an isolated system with the total energy $U$, the volume $V$ and $N$ particles; where $U$, $V$ and $N$, all have fixed values. Let $\Omega(U,V,N)$ be the number of different microstates which are consistent with this macro-state. To calculate $\Omega$ it is convenient to initially consider the total phase space volume as
\begin{equation}
\omega(U,V,N)=\frac{1}{h^{(3N)}}\int_{\sum E_i\leq U}\frac{d^{(3N)}xd^{(3N)}p}{\prod_{i=1}^N(1+\lambda^2E_{i}^2)^3}\hspace{.3 cm},
\end{equation}
where $E_i$ is the energy of the $i$th particle. Then we have
\begin{equation}
\Omega(U,V,N)=\frac{1}{N!}\frac{\partial \omega}{\partial U}\hspace{.3 cm},
\end{equation}
where $\frac{1}{N!}$ is the Gibbs correction factor. To show the existent of negative temperatures we do not need to calculate the above quantities explicitly. It is only sufficient to show that $\Omega(U,V,N)$ is not a monotonically increasing function of $U$ because the entropy of the system is directly proportional to $\ln \Omega$ and, as mentioned before, a system can have a negative temperature whenever the entropy is not restricted to a monotonically increasing function of the internal energy. It is easy to see that
\begin{equation}
\lim_{U\rightarrow\infty}\omega(U,V,N)=(\int_{0}^{\infty}\frac{\mathcal{C}E^2dE
}{h^{3}(1+\lambda^2 E ^2)^{3}})^N<\infty.
\end{equation}
 This means that the total number of accessible states is finite. Therefore, the limit of $\Omega(U,V,N)$ while $U$ tends to infinity is zero. Thus $\Omega(U,V,N)$ can not be a monotonically increasing function of $U$. We should note here that to avoid negative entropies and to preserve the third law of thermodynamics we have to set a cutoff on $U$ wherein the value of $\Omega(U,V,N)$ becomes one.

Now to calculate the thermodynamical quantities for negative temperatures by means of the grand canonical ensemble we should set a cutoff on energy in the integral
form of the partition function (\ref{5}). Obviously, the thermodynamical quantities depend on such a cutoff. But, as mentioned earlier, for a cutoff which is large enough the pervious calculations of the thermodynamical quantities in the positive temperature region are approximately valid. For example, the dispersion relation (\ref{4a}) may yield $E_c\sim\frac{L}{\hbar m}\lambda^{-3}$.
It is easy to show that by putting this cutoff in the integral
form of the partition function (\ref{5}), for non-negative
temperatures since $\lambda$ is very small, we have
\begin{equation}
\ln \mathcal{Z}_{c}(\beta)\simeq\ln \mathcal{Z}(\beta),\label{15}
\end{equation}
where $\mathcal{Z}_{c}$ denotes the partition function with
cutoff. It means that equations (\ref{8}) and (\ref{9}) are approximately valid for positive temperatures. But there is an important difference between two partition
functions. In the partition function $\mathcal{Z}_{c}$, it is
possible to have a negative value for $\beta$. A simple calculation shows
that the internal energy is always a monotonically decreasing
function of $\beta$ and the entropy is a monotonically decreasing
function only for positive values of $\beta$ and is a monotonically
increasing function for negative values as expected from the systems
allowing negative temperatures.

Having relation (\ref{15}) one can deduce that the relations
(\ref{11}) and (\ref{12}) approximately determine the thermodynamic
properties of the system for $\beta\simeq 0$. However,
it is important to note that this remains valid for both $\beta\gtrsim 0$
and  $\beta\lesssim 0$ since $\ln \mathcal{Z}_{c}$ is an analytic function of $\beta$.
We can also determine the energy density and the pressure of the
system for $\beta\simeq 0$:
\begin{equation}
\rho(x)\simeq\frac{U(x)}{V}\simeq
\frac{4\pi}{h^3\lambda^4}(\frac{1}{4}-\frac{3\pi }{16}x),\label{16}
\end{equation}
\begin{equation}
\mathcal{P}(x)\simeq\frac{4\pi}{h^3\lambda^4}(\frac{\pi}{16}\frac{1}{x}-\frac{1}{4}+\frac{3\pi}{32}x).\label{17}
\end{equation}
The energy density is always positive whether or not the temperature is positive. But, the pressure is positive only if the temperature is positive and is negative if the temperature is negative. Equation (\ref{17}) is valid only for $\beta\simeq 0$
but from equation (\ref{4}) it is easy to see that the pressure is
always negative for all negative temperatures. This is a novel and
interesting consequence of the generalized uncertainty principle. According to equation (\ref{4a}), one can see that if the length of the box $L$ increases, the energy of each level decreases. Then the question arises as how negative pressures can occur, whereas the energy of each level always decreases with increasing the volume of the system. It should be noted that increasing the volume of the system, however, increases the total number of accessible states and at negative temperatures because of the Boltzmann factor $e^{-\beta E_n}$, it causes the density of occupied states to increase near the cutoff. Thus, increasing the volume of the system (with a constant entropy) can cause an increase in the total internal energy which means that the pressure is negative.

\section{Removing the big bang singularity}

To investigate the effects of the generalized uncertainty principle (GUP) in the early universe two things have to be taken into account. First, we have to consider the change in the equations of state which is the result of the effect of the GUP on the behavior of the matter. Second, from equation (\ref{3b}) one can see that the GUP influences the structure of space-time because the space coordinates no longer remain commuting variables. Actually, in the high energy regime where the effect of the GUP becomes important, the structure and the geometry of space-time can not be described by a classical manifold and the Einstein field equations. Thus, we need a complete quantum theory of gravity. However, in the absence of such theory, one can effectively consider the effect of the GUP as a modification in the Einstein field equations. Therefore, the GUP assumption can modify the Friedmann equations \citep{Zhu09, Vakili08, Majumder11, Bina08, Barbosa04}. One can consider this modification as an effective consequence of the GUP on the dynamics of the universe.

 In this section, at the beginning, only the change in the equations of state is taken into account. In other words, for simplicity we naively assume the Friedmann equations are valid in the presence of the GUP. We then show that the consideration of the modified equation of state can alone remove the big bang singularity which usually appears in the solutions of the Friedmann equations in conventional cases. Finally, it will be shown that this result is also valid when we consider the modified Friedmann equations as the effect of the GUP on geometry.

Let us consider a reversible evolution with the case of zero spatial
curvature. Then the Friedmann equations may be written
as\footnote{We set $8\pi G=1$.}
\begin{equation}
(\frac{\dot{a}}{a})^2=\frac{1}{3}\rho,\label{18}
\end{equation}
\begin{equation}
\frac{\ddot{a}}{a}=-\frac{1}{6}(\rho+3\mathcal{P}),\label{19}
\end{equation}
where $a$ is the scale factor and dots denote derivatives with
respect to proper time. Instead of the second equation one may use
the conservation equation
\begin{equation}
\dot{\rho}+3\frac{\dot{a}}{a}(\rho+\mathcal{P})=0.\label{20}
\end{equation}
For an expanding universe, using equation (\ref{18}) this leads to
\begin{equation}
\dot{\rho}+\sqrt{3\rho}(\rho+\mathcal{P})=0.\label{21}
\end{equation}
If we substitute the equations of state (\ref{16}) and (\ref{17}) in
this equation then we have
\begin{equation}
\dot{x}=\sqrt{\frac{\pi}{3h^3\lambda^4}}(1-\frac{3\pi}{4}x)^{\frac{1}{2}}(\frac{1}{x}-\frac{3}{2}x).\label{22}
\end{equation}
Note that this equation is valid only for $x\simeq 0$. Also it is
easy to see that these equations yield
\begin{equation}
a^3(1-\frac{3}{2}x^2)=\textrm{constant}.\label{23}
\end{equation}
This relation means that the entropy is constant (see equation
(\ref{12})) which is consistent with the assumption of
reversibility. The most interesting result of this relation is that
the minimum value of the scale factor $a$ is greater than zero, in
contrast to the conventional case. The minimum of the scale factor
occurs at infinite temperature or $x=0$, like the conventional case.
But the difference is that the term $S/a^3$ does not tend to zero as
$x\rightarrow 0$. This result is true even when we assume negative
temperatures and also in the case of non-zero spatial curvature. Actually, this is a consequence of the fact that the
entropy in a constant volume is finite at $x=0$. Note that this result is not a trivial consequence of the existence of a non-zero minimal uncertainty in position measurements. For example, one can
see that the generalized Heisenberg algebra (\ref{2}) cannot
remove the big bang singularity \citep{Alexander01}.

  As it was mentioned earlier, the existence of a non-zero minimal uncertainty in position measurements not only changes the equations of state of matter but also modifies the Friedmann equations. This modification can be considered as an effective correction because in case of modified Friedmann equations we assume the background manifold as a classical manifold. In order to modify the Friedmann equations due to the effect of the GUP one can employ different approaches. For example, in \citep{Vakili08} the author assumes that the GUP deforms its corresponding Poisson algebra between the scale factor and its momentum conjugate in the Friedmann-Robertson-Walker (FRW) universe. It has been then shown that this affects the Friedmann equations. Another approach can be based on the fact that the GUP applies a correction on the conventional entropy-area relation of the apparent horizon of the FRW universe \citep{Zhu09}. In a FRW space-time, assigning the temperature
  \begin{equation}
T=\frac{1}{2\pi\tilde{r}_A}
\end{equation}
and the entropy
\begin{equation}
S=\frac{A}{4G}\label{24}
\end{equation}
to the apparent horizon and using the Clausius relation $\delta Q=TdS$ , one can obtain the Friedmann equations \citep{Cai05}. Here $G$, $A$ and $\tilde{r}_A$ are the gravitational constant, the area and the radius of the apparent horizon, respectively. Therefore, applying this method to the corrected entropy-area relation obtained by the GUP assumption, one can modify the Friedmann equations \citep{Zhu09}. Now, we assume that the modified Friedmann equations obtained in \citep{Zhu09} can effectively describe the effect of the GUP on the dynamics of the FRW universe. We do not deal with the details of the calculation and only employ the results obtained in \citep{Zhu09}. For a $(3+1)$-dimensional FRW universe whose line element is written as
\begin{equation}
ds^2=dt^2-a^2(\frac{dr^2}{1-kr^2}+r^2 d\Omega^2),\label{24a}
\end{equation}
the modified Friedmann equations is given by
\begin{equation}
(\dot{H}-\frac{k}{a})f(A)=-4\pi G(\rho+\mathcal{P}),   \label{25}
\end{equation}
\begin{equation}
\frac{8\pi G}{3}\rho=-4\pi\int f(A)(\frac{1}{A})^2 dA,     \label{26}
\end{equation}
where $A=4\pi\tilde{r}_{A}^{2}$ , $H$ is the Hubble parameter and $f(A)$ is a function of $A$ which is determined by the generalized uncertainty relation. Using the line element (\ref{24a}), the radius of the apparent horizon becomes
\begin{equation}
\tilde{r}_{A}=\frac{1}{\sqrt{H^2+k/a^2}}. \label{27a}
\end{equation}
For the generalized uncertainty relation (\ref{3}) the function $f(A)$ is given by
\begin{equation}
f(A)=\frac{A}{2\pi\hbar\lambda^2}(1-\sqrt{1-\frac{4\pi\hbar^2\lambda^2}{A}}). \label{27}
\end{equation}
It is easy to see that when the parameter $\lambda$ tends to zero the modified Friedmann equations reduce to the conventional Friedmann equations. Now, we claim that by applying the modified equations of state (\ref{16}) and (\ref{17}) to the modified Friedmann equations, we again obtain the relation (\ref{23}). Note that the modified Friedmann equations with the relation (\ref{27}) imply
\begin{equation}
\dot{\rho}+3H(\rho+\mathcal{P})=0,
\end{equation}
which is the continuity equation of the perfect fluid. This equation can be re-expressed as
\begin{equation}
d(\rho V)+\mathcal{P}dV=0
\end{equation}
where $V$ is the volume of the universe and is proportional to $a^3$. comparing this equation with the first law of thermodynamics, one can deduce that the entropy of the perfect fluid is a constant quantity which yields the relation (\ref{23}). As said before, this relation implies that there is no singularity in the solution of the dynamical equations. Therefore, in accordance with the above calculations one can argue that this result is independent of the dynamical equations of the universe. In other words, the reversibility assumption which implies a constant entropy with the modified equations of state is enough to avoid the singularities.

 It is important to point out that since the GUP is a high energy effect and in this regime the effect of quantum gravity should not be forgotten, to investigate the ultimate fate of the big bang singularity a fully quantized theory such as quantum cosmology is required. But, this subject is out of the scope of this paper.

Another important thing to note here is that in the negative
temperature region there are some new solutions for the Friedmann equations and the modified Friedmann equations. For example, this makes it possible to have an expanding and accelerating solution for
equation (\ref{19}). It means that the negative temperature consideration generally changes the history of the universe.
Thus, the existence of a non-zero minimal uncertainty in position measurements with the assumption of negative temperatures might open new ways to study the early universe problems. The investigation of the effects of these assumptions on the dynamics of the universe remains a subject of further studies.

\acknowledgments
The author is very thankful to Prof. F. Arash and Dr. G. R. Jafari for helpful and enlightening discussions. This work was supported by Shahid Rajaee Teacher Training University under contract number 36503.

%%%%%%%%%%%%%%%%%%%%%%%%%%%%%%%%%%%%%%%%%%%%%%%%%%%%%%%%%%%%%%%%%%%%%%%%%%%

\makeatletter
\let\clear@thebibliography@page=\relax
\makeatother


\begin{thebibliography}

\bibitem[Alexander \& Magueijo (2001)]{Alexander01} Alexander, S. \& Magueijo, J., 2001, arXiv:hep-th/0104093.

\bibitem[Amati, Cialfaloni, \& Veneziano (1989)]{Amati89}  Amati, D., Cialfaloni, M., Veneziano, G., 1989, Phys. Lett. B216, 41.

\bibitem[Ashtekar, Bojowald,\& Lewandowski (2003)]{Ashtekar03} Ashtekar, A., Bojowald, M., \& Lewandowski, J., 2003, Adv. Theor. Math. Phys. 7, 233.

\bibitem[Barbosa \& Pinto-Neto (2004)]{Barbosa04} Barbosa, G. D., \& Pinto-Neto, N. , 2004, Phys. Rev. D 70, 103512.

\bibitem[Bina, Atazadeh, \& Jalalzadeh (2008)]{Bina08} Bina, A., Atazadeh, K., \&  Jalalzadeh, S., 2008,  Int. J. Theor. Phys. 47, 1354.

\bibitem[Bojowald (2001)]{Bojowald01} Bojowald, M., 2001, Phys. Rev. Lett. 86, 5227, (arXiv:gr-qc/0102069).

\bibitem[Borde \& Vilenkin (1994)]{Borde94} Borde, A., \& Vilenkin, A., 1994, Phys. Rev. Lett. 72, 3305.

\bibitem[Brout et al. (1999)]{Brout99} Brout, R., Gabriel, Cl., Lubo, M., \& Spindel, Ph., 1999, Phys. Rev. D 59, 044005, (arXiv:hep-th/9807063).

\bibitem[Cai \& Kim (2005)]{Cai05} Cai, R. G., \& Kim, S. P., 2005, JHEP 02, 050.

\bibitem[Chang et al. (2002)]{Chang02} Chang, L. N., Minic, D., Okamura, N., \&  Takeuchi, T., 2002, Phys. Rev. D 65, 125028, (arXiv:hep-th/0201017).

\bibitem[Garay (1995)]{Garay95} Garay, L. J., 1995, Int. J. Mod. Phys. A10, 145.

\bibitem[Gasperini \& Veneziano (1993)]{Gasperini93} Gasperini, M., \& Veneziano, G., 1993, Astroparticle Phys. 1, 317.

\bibitem[Greiner, Neise, \& Stocker (1995)]{Greiner95} Greiner, W., Neise, L., \& Stocker, H., 1995, Thermodynamics and Statistical Mechanics, Springer-Verlag.
    
\bibitem[Hawking \& Ellis (1973)]{Hawking73} Hawking, S. W., \& Ellis, G. F. R., 1973, The large scale structure of space-time, Cambridge University Press.
    
\bibitem[Jaeckel \& Reynaud (1994)]{Jaeckel94}  Jaeckel, M.-T., \&  Reynaud, S. , 1994, Phys. Lett. A185, 143.

\bibitem[Kawai et al. (2006)]{Kawai06} Kawai, S., Keski-Vakkuri, E., Leigh, R. G., \& Nowling, S., 2006, Phys. Rev. Lett. 96, 031301.

\bibitem[Kempf, Mangano, \& Mann (1995)]{Kempf95} Kempf, A., Mangano, G., \& Mann, R. B., 1995, Phys. Rev. D 52, 1108, (arXiv:hep-th/9412167).

\bibitem[Larsen \& Wilczek (1997)]{Larsen97} Larsen, F., \& Wilczek, F., 1997, Phys. Rev. D 55, 4591.

\bibitem[Lubo (2000a)]{Lubo00a} Lubo, M., 2000a, Phys. Rev. D 61, 124009, (arXiv:hep-th/9911191).

\bibitem[Lubo (2000b)]{Lubo00b} Lubo, M., 2000b, arXiv: hep-th/0009162.

\bibitem[Maggiore (1993a)]{Maggiore93a} Maggiore, M., 1993a, Phys. Lett. B 304, 65, (arXiv:hep-th/9301067).

\bibitem[Maggiore (1993b)]{Maggiore93b} Maggiore, M., 1993b, Phys. Lett. B 319, 83, (arXiv:hep-th/9309034).

\bibitem[Maggiore (1994)]{Maggiore94} Maggiore, M., 1994, Phys. Rev. D 49, 5182, (arXiv:hep-th/9305163).

\bibitem[Majumder (2011)]{Majumder11} Majumder, B., 2011, Astrophys. Space Sci. 336,  331.

\bibitem[Pathria (1996)]{Pathria96} Pathria, R. K., 1996, Statistical Mechanics-2nd ed.\\ Butterworth-Heinmann.

\bibitem[Rama (2001)]{Rama01} Rama, S. K., 2001, Phys. Lett. B 519, 103.

\bibitem[Ramsey (1956)]{Ramsey56} Ramsey, N. F., 1956, Phys. Rev. 103, 20.

\bibitem[Reif (1985)]{Reif85} Reif, F., 1985, Fundamentals of Statistical and Thermal Physics, McGRAW-HILL.

\bibitem[Vakili (2008)]{Vakili08} Vakili, B., 2008, Phys. Rev. D 77, 044023.

\bibitem[Zhu, Ren, \& Li (2009)]{Zhu09} Zhu, T., Ren, J.-R., \&  Li, M.-F., 2009, Phys. Lett. B 674, 204.
\end{thebibliography}
\end{document}